\documentclass[manuscript]{aastex631}

\newcommand{\SFR}{$\rm \Delta \Sigma_{SFR}$}
\newcommand{\age}{$\rm Age_L$}
\newcommand{\grad}{$\nabla_{1R_e}$}

\usepackage{amsmath}

\begin{document}

\title{Resolved star formation in TNG100 central and satellite galaxies}

\author[0000-0001-6928-4345]{Bryanne McDonough}
\affiliation{Institute for Astrophysical Research, Boston University, 725 Commonwealth Ave, Boston, MA}https://www.overleaf.com/project/6397926f74163487fc461872

\author[0000-0002-0212-4563]{Olivia Curtis}
\affiliation{Institute for Astrophysical Research, Boston University, 725 Commonwealth Ave, Boston, MA}

\author[0000-0001-7917-7623]{Tereasa G. Brainerd}
\affiliation{Institute for Astrophysical Research, Boston University, 725 Commonwealth Ave, Boston, MA}



\begin{abstract}

Recent cosmological hydrodynamical simulations have produced populations of numerical galaxies whose global star-forming properties are in good agreement with those of observed galaxies. 
Proper modeling of energetic feedback from supernovae and active galactic nuclei is critical to the ability of simulations to reproduce observed galaxy properties and, historically, such modelling has proven to be a challenge. 
Here, we analyze local properties of central and satellite galaxies in the $z=0$ snapshot of the TNG100 simulation as a test of feedback models. 
We generate a face-on projection of stellar particles in TNG100 galaxies, from which we demonstrate the existence of a resolved star-forming main sequence ($\rm \Sigma_{SFR}$--$\Sigma_*$ relation) with a slope and normalization that is in reasonable agreement with previous studies. 
We also present radial profiles of various galaxy populations for two parameters: the distance from the resolved main sequence line (\SFR) and the luminosity-weighted stellar age (\age). 
We find that, on average, high-mass central and satellite galaxies quench from the inside-out, while low-mass central and satellite galaxies have similar, flatter profiles. 
Overall, we find that, with the exception of the starburst population, the TNG100 feedback models yield simulated galaxies whose radial distributions of \age{} and \SFR{} agree with those of observed galaxies.
\end{abstract}



\section{Introduction} \label{sec:intro}
Over the last few decades, numerical simulations have become an important tool for understanding galaxy formation and evolution. Early on it became apparent that it was necessary to include feedback processes within hydrodyamical simulations in order to prevent early, overly-efficient star formation (known as the over-cooling problem; see, e.g.,  \citealt{Naabreview} for a review). 
In recent simulations, feedback from supernovae and Active Galactic Nuclei (AGN) have been implemented to address this problem \citep[e.g.,][]{Weinberger17}. The feedback models in the Illustris-1 simulation \citep{Vogelsberger14,Nelson15} resulted in a lack of massive galaxies with low levels of star formation at low redshifts \citep{Genel14} because star formation was not being shut down efficiently within these massive systems. The successor suite of simulations, IllustrisTNG (hereafter, TNG), made use of an updated feedback model \citep{Weinberger17}, resulting in a numerical galaxy population with optical colors that agree well with those of observed, low redshift galaxies \citep{Nelson2018}. 

In observations \citep[e.g.,][]{Brinchmann04, Cano-Diaz16,Renzini15,Speagle_2014} and in TNG \citep{Donnari19}, a distinct correlation between the global stellar mass and star formation rate (SFR) exists for star forming galaxies. This is known as the star forming main sequence (hereafter, the ``main sequence'') and it is in place from $z\sim6$ to the present day \citep[e.g.,][]{Speagle_2014,Popesso22}. 
In parallel to the trend in global stellar mass and SFR for observed galaxies, there is also a trend in the resolved stellar mass surface density ($\Sigma_*$) and SFR density ($\Sigma_{\mathrm{SFR}}$), commonly referred to as the resolved star forming main sequence (rSFMS). The rSFMS has been identified in photometric observations \citep[e.g.,][]{Maragkoudakis16,Abdurro'uf17} and via integral field spectroscopy \citep[IFS, e.g.,][]{Cano-Diaz16,Ellison18,Liu_2018,Medling18,Erroz-Ferrer,Bluck2020a}. 

\cite{Trayford19} identified the resolved main sequence in the EAGLE simulation \citep{Crain15,Schaye15} and \cite{Hani20} identified the resolved main sequence in the FIRE-2 \citep{Hopkins18} zoom-in simulations of Milky Way-like galaxies. Further, \cite{Hani20} showed that the slope of the rSFMS is affected by the spatial resolution of the simulation (i.e., due to the clumpy nature of star formation), which is especially important to consider when comparing slopes of the rSFMS across various studies. 

Radial profiles of properties that relate to star formation can be particularly informative when investigating how and why galaxies transition away from the global SFMS. Previous studies \citep{GonzalezDelgado16,Ellison18,Spindler18,Bluck2020b} have shown that the specific star formation rate (sSFR), or the offset from the rSFMS ($\Delta \Sigma_{\mathrm{SFR}}$), is suppressed toward the centers of massive ($\log_{10} M_*/M_\odot\gtrsim 10$) galaxies, with relatively more star formation occurring in the outskirts of these objects. This has been labelled ``inside-out'' quenching, since the galaxies quench first in the center, with the quenching proceeding outwards over time.

\cite{Bluck2020b} expanded on earlier results by using both emission lines and an empirical relationship to obtain SFRs, resulting in a larger sample of spaxels (i.e., spectral pixels from IFS observations), particularly those with lower levels of star formation. Their results largely agree with previous studies, finding that most green valley galaxies quench from the inside-out. However, they found that satellite galaxies in the green valley exhibit flatter $\Delta \Sigma_{\mathrm{SFR}}$ profiles, indicating that inside-out quenching primarily occurs in massive central galaxies. Additionally, low-mass ($\log_{10} M_*/M_\odot < 10.0$) galaxies did not show evidence for inside-out quenching, and \cite{Bluck2020b} did not rule out the possibility of outside-in quenching for these systems. 

\cite{Nelson21} examined radial profiles of sSFR in TNG50, the highest resolution simulation in the IllustrisTNG suite \citep{TNG50_Nelson,TNG50_Pillepich}. Using instantaneous SFRs (computed via summing over the SFRs of individual gas cells), \cite{Nelson21} compared radial profiles of simulated galaxies to 3D-HST observations and found that the profiles agree well, with both showing inside-out quenching for massive systems ($\log_{10} M_*/M_\odot >10.5$). Additionally, \cite{Nelson21} showed that galaxies in the predecessor simulation, Illustris-1, did not display signatures of inside-out quenching. Since the major difference between the physical models in the two simulations is the way in which AGN feedback was implemented, \cite{Nelson21} concluded that quenching in the central regions of massive TNG galaxies is associated with central AGN.

While the radial profiles of star formation in TNG50 agree with observations, the results of \cite{Bluck2020b} indicate that quenching may proceed differently in central galaxies, as compared to satellite galaxies. Additionally, \cite{Bluck2020b} found that the degree of quenching scales with galaxy mass. This motivates us to perform a similar investigation using the larger volume TNG100 simulation. While lower in resolution than TNG50, the larger volume of TNG100 provides a larger sample size, allowing statistical analyses of radial profiles for different populations of galaxies to be carried out.

While simulations are now reproducing global properties of galaxies reasonably well, how and where feedback should be implemented in the models is still being debated. Therefore, an important test of feedback models is the degree to which local properties of simulated galaxies agree with observational results from IFS surveys. Here, we examine radial profiles of star formation and stellar age for various TNG100 galaxy populations and identify whether, and where, the simulation agrees with observations. Such a test could reveal either areas where the physical models need improvement, or demonstrate that radial profiles of star-forming properties in TNG100 galaxies are in good agreement with observations. Showing the latter would motivate future studies of the dependency of star formation radial profiles on various intrinsic and environmental properties, some of which can only be probed easily in simulation space.

Radial profiles of star formation are also of interest for starburst galaxies, which are located well above the main sequence. For example, \cite{Ellison18} and \cite{Bluck2020b} found that, on average, starbursts have the star bursting occurring in their centers. However, analogous studies may not yet be feasible with cosmological simulations. For example, \cite{Sparre16} used zoom-in simulations of Milky Way-like galaxies to show that the peak of star formation in merger-induced star bursts in the Illustris-1 simulation is limited by spatial resolution. They attribute this to higher resolution simulations having a better sampling of smaller, denser regions. Here we also investigate radial profiles of starburst galaxies, but since the spatial resolution of Illustris-1 and TNG100 are identical, the profiles of TNG starbursts that we present should be interpreted with caution.

The paper is organized as follows. In Section \ref{sec:data}, we describe TNG100, along with the data and supplementary catalogs that we use. In Section \ref{sec:methods}, we describe the ways in which time-averaged star formation rates were obtained, a global star-forming main sequence was constructed, and maps of resolved star formation were generated.  In addition, we present a resolved star-forming main sequence for TNG100 galaxies. In Section \ref{sec:results}, we present results for radial profiles of two parameters: luminosity-weighted stellar age (\age) and offset from the resolved main sequence (\SFR). In Section \ref{sec:discussion}, we compare our results to those of observational surveys and other simulations. A summary of our results and main conclusions are presented in Section \ref{sec:conclusion}.

\section{Data} \label{sec:data}
Our galaxy sample is drawn from the $z = 0$ snapshot of the
TNG100-1 simulation (hereafter, TNG100).  The IllustrisTNG project is a publicly available suite of $\Lambda$CDM magnetohydrodynamical simulations \citep{Nelson2018,2018MNRAS.475..648P,2018MNRAS.475..676S,2018MNRAS.477.1206N,2018MNRAS.480.5113M,Nelson2019a} that adopted a \citet{Planck2016} cosmology with the following parameter values:
$\Omega_{\Lambda,0}=0.6911$, $\Omega_{m,0}=0.3089$, $\Omega_{b,0}=0.0486$, $\sigma_8=0.8159$, $n_s=0.9667$, and $h=0.6774$.  
The co-moving volume of TNG100 is $75^3 h^{-3}\rm{Mpc}^3$, and it contains $1820^3$ dark matter particles and $1820^3$ gas particles at the start of the simulation.
Here, TNG100 was chosen because it offers good resolution of relatively low-mass ($\sim10^{9} M_\odot$) satellite galaxies while also allowing investigation of a substantial number of central and satellite galaxies. 
Excluding subhalos that are flagged as being non-cosmological in origin, there are $\sim 60,000$ luminous TNG100 galaxies with $M_r < -14.5$ at $z = 0$.

From the \texttt{SubFind} catalog, we take the galaxy stellar masses, $M_*$, to be the total mass of all stellar particles that are bound to each galaxy. The minimum mass of a stellar particle in TNG100 is $\sim 7\times10^6 M_\odot$, which effectively sets a minimum detectable rate for star formation, since at least one stellar particle must be generated in the region of interest for an SFR to be calculated. 

In addition to the main simulation results, we also make use of supplementary catalogs for the TNG100 galaxies. To aid in comparison to results from observational studies, we use the Stellar Projected Sizes catalog from \citet{Genel2018}. From this catalog, we use the $r$-band half-light radius from the face-on projection of each galaxy, which we will refer to as the effective radius, $R_e$. For magnitudes, we use the SDSS Photometry, Colors, and Mock Fiber Spectra catalog from \citet{Nelson2018}, which includes the effects of dust obscuration.

To ensure sufficient mass resolution, we limit our sample to galaxies containing at least $1,000$ stellar particles within $2R_e$, which corresponds roughly to a minimum stellar mass of $10^{9} \: M_\odot$. Additionally, to ensure sufficient spatial resolution, we restrict the sample to galaxies with $R_e > 4$ kpc (i.e., five times the gravitational softening radius).  There are a total of $6197$ TNG100 galaxies that meet these criteria, consisting of $4605$ central galaxies and $1592$ satellites. Here we define a central galaxy to be the most massive galaxy in a group identified by the friends-of-friends algorithm, and we consider any other galaxy in the group to be a satellite.  

\section{Methods} \label{sec:methods}
\subsection{Comparing Observational and Simulated SFRs} \label{sec:3.1}

Obtaining star formation rates, whether in simulations or observations, can be challenging. In observations, star formation is largely hidden in dense molecular clouds and must be inferred based on the presence of various tracers. These tracers track star formation over different timescales and must be calibrated carefully. In simulations, SFRs can be calculated more directly, but resolution is a limiting factor and, when comparing results from observations to those of simulations, SFRs from simulations should be computed in a way that is most comparable to observations. Here, we construct our SFRs to roughly mimic those obtained from the Mapping Nearby Galaxies at Apache Point Observatory survey \citep[MaNGA, ][]{MaNGA} by \cite{Bluck2020a} and used by \cite{Bluck2020b}. The MaNGA survey obtained spatially resolved spectroscopic measurements of $\sim10,000$ nearby galaxies and \cite{Bluck2020a} derived parameter maps from the spectroscopic data cubes. 

\cite{Bluck2020a} used two common tracers of star formation: [1] the \cite{Kennicutt98} relation for the dust-corrected H$\alpha$ flux, where possible, and [2] an empirical calibration between the strength of the 4000 $\AA$ break (D4000) and sSFR. These two methods trace star formation over different timescales. The H$\alpha$ lines originate in H~II regions around young, massive stars with ages $\lesssim 20$ Myrs \citep{Kennicutt98}.The timescale for the D4000 estimator is more complicated since it is computed as a ratio between the emission at $4050$ -- $4250 \AA$ and the emission at $3750$ -- $3950 \AA$. The break is due to absorption lines of heavy elements causing a sudden change in opacity of the stellar atmospheres in stars older than $\sim100$ Myrs \citep{EGbook}. Stars younger than $\sim 100$ Myrs will still contribute significant light in the $3750$ -- $3950 \AA$ band \citep{Caplar}. Therefore, the D4000 estimator roughly traces star formation over the last $\sim 100$ Myrs, although it can be affected by contamination from older stellar populations. For levels of star formation that were undetectable with sufficient signal-to-noise, \cite{Bluck2020a} assigned a fixed value of $\rm sSFR =10^{-12} yr^{-1}$. Due to this, \cite{Bluck2020b} note that low values of SFR should be treated as upper limits. 

While observational tracers of SFR frequently overestimate the SFR by 25\% -- 65\% \citep[e.g.,][]{Boquien}, several factors can affect SFRs recovered from simulated galaxies, including mass resolution, averaging timescales, and spatial apertures \citep[see][for a discussion of these factors in the TNG simulations]{Donnari19, erratum}.

Here, we use time-averaged SFRs, rather than instantaneous values. These are lower limits, however, because small amounts of star formation in quenched galaxies are not resolved. Therefore, for undetectably low levels of star formation, we adopt the same fixed sSFR as \cite{Bluck2020a}, meaning that our lowest levels of star formation are also upper limits. The minimum time-averaged SFR is pre-determined by the minimum baryonic particle mass ($7\times10^5 \: \mathrm{M}_\odot$) and the timescale over which it is averaged. 


\subsection{Global Star Formation} \label{sec:gsf}

We will use two timescales (20 Myr and 100 Myr), over which we find the average SFR by identifying the initial mass of all particles that are bound to the subhalo, located within $2 R_e$ of the center, and formed within the last $X$~Myr, where $X = 20$ or 100. The 20 and 100~Myr timescales are chosen to roughly correspond to the timescales of the H$\alpha$ and D4000 SFR tracers from \cite{Bluck2020a}. We compute global SFRs over a 20 Myr timescale whenever possible (i.e., galaxies in which at least one stellar particle formed in the past 20~Myr), otherwise we use the 100 Myr timescale. Of the $6197$ galaxies in our sample, SFRs for $4868$ galaxies can be computed with the 20 Myr timescale, SFRs for $289$ galaxies are computed with the 100 Myr timescale, but $1041$ galaxies have not formed any new stellar particles in the last 100 Myr. For these galaxies we introduce a fixed sSFR of $10^{-12} \rm{yr}^{-1}$ with random scatter ($\sigma = 0.1$ dex). This is done for two reasons: [1] it insures that quenched galaxies for which low levels of star formation are unresolved have small, but non-zero sSFR values, and [2] it makes our analysis of the quenched galaxies consistent with the approach used by \cite{Bluck2020b}. We note that a TNG supplementary catalog of time-averaged SFRs exists \citep{2018MNRAS.475..648P, Donnari19}; however, since the catalog does not include a 20~Myr timescale, we choose to calculate all of the SFRs for consistency with \cite{Bluck2020a}. 

Due to the limited resolution of the simulation, each SFR timescale has a fixed lower limit, determined by the timescale itself and the minimum baryonic particle mass. For the 20 Myr and 100 Myr timescales, the smallest non-vanishing SFR values are $\mathrm{log_{10}(SFR}_{\mathrm{min}} / [\mathrm{M}_\odot \rm{yr}^{-1} ]) \: \simeq \: 1.46,$ and $-2.15$, respectively. 

\begin{figure}[tp]
	\centering
	
	\includegraphics[width=\linewidth]{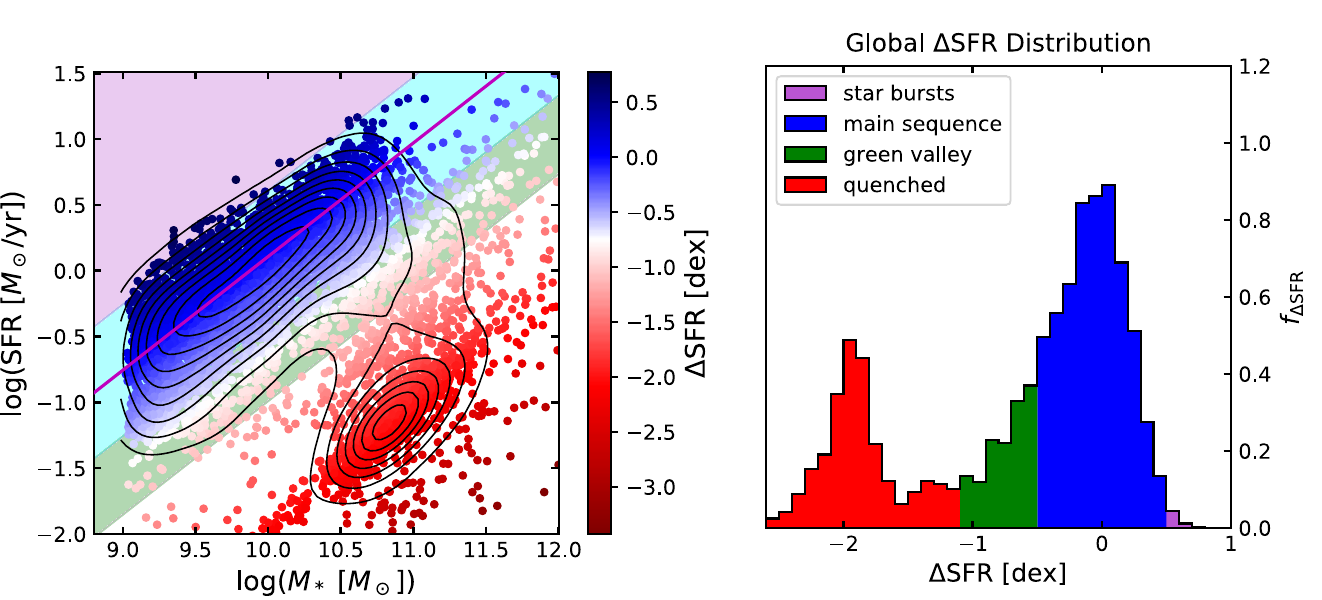}

    \caption{{\it Left:} Global star forming main sequence for resolved TNG100 galaxies, where global star formation rate is plotted against stellar mass (measured as the sum of all stellar particles bound to the subhalo). Galaxies that have not experienced any new star formation in the last 100 Myr are assigned a fixed value of $\rm sSFR=10^{-12} yr^{-1}$ with some scatter. Point color is determined by $\Delta$SFR, the mean logarithmic distance to the global main sequence (magenta line). Shaded areas indicate the classification of galaxies in that region (starburst: purple, main sequence: blue, green valley: green, quenched: white). {\it Right:} Distribution of $\Delta$SFR values. Colored regions indicate where galaxies are classified as starburst (purple), main sequence (blue), green valley (green), and quenched (red).}
    \label{fig:glMS}
    
\end{figure}

In Figure \ref{fig:glMS} (left), we show the main sequence (SFR--$M_*$ relation) for TNG100 galaxies. Point colors are based on their distance from the main sequence line ($\Delta$SFR), which is shown in magenta. Contours indicate the density of points. Like observed galaxies, there is a bimodal distribution in the SFR--$M_\ast$ relation for TNG100 galaxies. However, the red peak for TNG100 galaxies is largely comprised of galaxies with SFRs that were assigned the fixed lower sSFR value. Without the assignment of the fixed sSFR value, there is no discernible red peak, since many of the quenched galaxies have a time-averaged SFR of 0. 

In fitting the main sequence, we follow \citet{Renzini15} and define the main sequence ridge line using the full sample of galaxies (i.e., both star-forming and quenched galaxies). In practice, this means we identify the mode of the distribution in stellar mass bins of $\log_{10} M_*/M_\odot \sim 0.06 M_\odot$ and then perform a least-squares fit. The fit is performed in the range $9.0 \leq \log_{10}M_*/M_\odot \leq 10.4$. The limited volume of TNG100 means that sampling of the most massive galaxies is incomplete, so we choose to fit within this lower mass range for which the sample is not limited by volume. Additionally, there is some debate over whether the main sequence continues linearly at high masses (e.g., \citealt{Popesso18,Whitaker_2015,Renzini15}). For stellar masses $> 10^9 M_\odot$, however, \cite{Sparre15} find that the normalization of the main sequence in the original Illustris simulation is well converged.

From Figure \ref{fig:glMS}, the main sequence is given by:
\begin{equation}
    \centering
    \log_{10}\mathrm{SFR_{MS}} = (0.86 \pm 0.05)\times \log_{10}(M_*/M_\odot) - (8.5 \pm 0.5),
    \label{eq:SFMS}
\end{equation}
where the errors are the standard deviations of the fit. The quantity $\Delta$SFR then is computed as:
\begin{equation}
    \centering
    \Delta\mathrm{SFR} = \log_{10}(\mathrm{SFR}) - \log_{10}(\mathrm{SFR_{MS}}),
\end{equation}
which separates star-forming and quenched galaxies, regardless of stellar mass. The distribution of $\Delta$SFR values is shown in Figure \ref{fig:glMS} (right). There is a clear bimodal distribution of $\Delta$SFR values, although the red peak is largely artificial, since it corresponds to the fixed sSFR we adopted for galaxies with unresolved, low levels of star formation. The blue peak is located at $\rm \Delta SFR \sim 0$, as expected from the definition of the main sequence line.  

The $\Delta$SFR parameter allows for the classification of galaxies on the basis of their distance to the star-forming main sequence and we use this to define four categories of galaxies: those far above, on, below, or far below the main sequence line. Table \ref{tab:class} summarizes the definitions of each galaxy class (starburst, main sequence, green valley, quenched) and lists the number of galaxies (total, central, and satellite) in each class. Our definitions for star-forming classes are adopted from \cite{Bluck2020b}.

\begin{deluxetable}{lcccc}
\tablecaption{Definitions for the four classes of galaxies based on $\Delta$SFR. \label{tab:class}}
\tablehead{
\colhead{Classification} &
\colhead{$\Delta$SFR} &
\colhead{$\mathrm{N_{gals}}$} &
\colhead{$\rm N_{centrals}$} &
\colhead{$\rm N_{satellites}$}
}
\startdata
Starbursts (SB) & $\Delta$SFR $> 0.5$ dex & 58 & 38 & 20 \\
Main sequence (MS) & $-0.5$ dex $< \Delta$SFR $< 0.5$ dex & 3624 & 2942 & 682 \\
Green valley (GV) & $-1.1$ dex $< \Delta$SFR $< -0.5$ dex & 829 & 636 & 193 \\
Quenched (Q) & $\Delta$SFR $< -1.1$ dex & 1687 & 989 & 698 \\
\enddata
\tablecomments{Here, $\rm N_{gals}$ is the total number of galaxies of a given classification, while $\rm N_{centrals}$ and $\rm N_{satellites}$ indicate, respectively, the number of central and satellite galaxies in each classification.}

\end{deluxetable}



\begin{deluxetable}{lccccc}
\tablecaption{Slopes and zero points for the $z=0$ global main sequence (Equation \ref{eq:SFMS}).  \label{tab:SFMS}}
\tablehead{
\colhead{Paper} &
\colhead{Data Source} &
\colhead{Data Type} &
\colhead{Fitting Technique} &
\colhead{Slope} &
\colhead{Zero Point}
}
\startdata
this work & TNG100 & simulation & ridge line & $0.86 \pm 0.05$ & $-8.5 \pm 0.5$ \\
\cite{erratum} & TNG300 & simulation & OLS & $0.80 \pm 0.01$ & $-8.15 \pm 0.11$ \\
\cite{Speagle_2014} & pre-2014 census & observations & OLS & $0.84 \pm 0.02$ & $-6.51 \pm 0.24$ \\
\cite{Renzini15} & SDSS DR7 & observations & ridge line & $0.76 \pm 0.01$ & $-7.64 \pm 0.02$ \\
\cite{Cano-Diaz16} & CALIFA & observations & OLS & $0.81 \pm 0.02$ & $-8.34 \pm 0.19$ \\
\enddata
\tablecomments{All investigations in this table adopted a linear definition of the main sequence. OLS - Ordinary Least Squares}
\end{deluxetable}

Table \ref{tab:SFMS} compares the MS slope we find for TNG100 galaxies to the MS slope found in other studies.  We will discuss this in the context of previous work in Section \ref{sec:compGMS} and here we simply note that, considering the various methods and assumptions that are made, the MS slope we find for TNG100 galaxies is in good agreement with results from previous studies of observed and simulated galaxies.



\subsection{Local Star Formation} \label{sec:lsf}
\begin{figure}[tp]
	\centering
    \includegraphics[width=\linewidth]{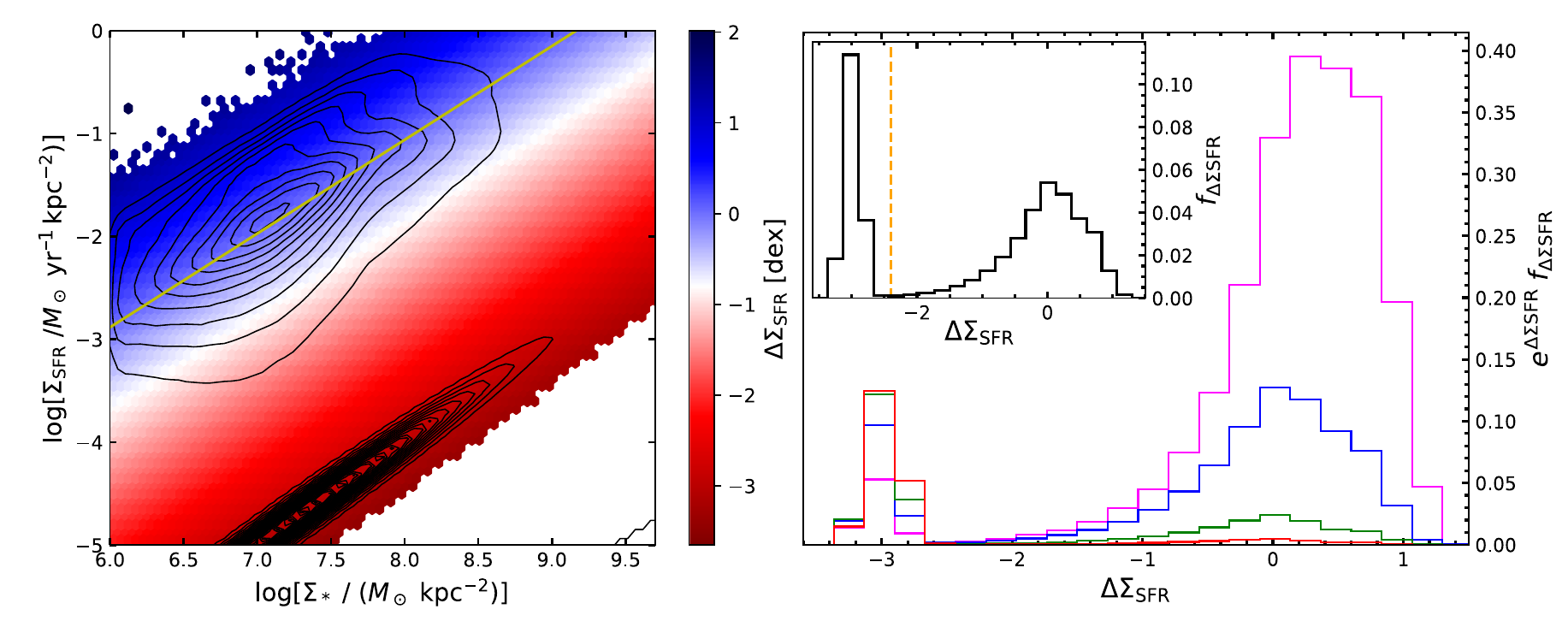}
    \caption{{\it Left:} Resolved star forming main sequence for ``spaxels'' from maps of TNG100 galaxies. Star formation rate surface density is plotted against stellar mass surface density for each spaxel, which has been hexagonally binned. Point color is determined by $\rm \Delta\Sigma_{SFR}$.  Yellow line:  the resolved main sequence.  {\it Right:} Distribution of $\rm \Delta\Sigma_{SFR}$ values, weighted by $\rm e^{\Delta \Sigma_{SFR}}$. Main figure: distribution of spaxels for each galaxy type (starbursts: pink, main sequence: blue, green valley: green, quenched: red). Inset: distribution of all spaxels from the entire sample. Orange dashed line: minimum of the distribution, below which a spaxel is considered to be quenched. }
    \label{fig:resMS}
\end{figure}

In analogy to the global main sequence, a relationship between $\Sigma_*$ and $\Sigma_{\mathrm{SFR}}$ for spaxels has been found in IFU observations. This has been referred to as the resolved main sequence. In this section, we discuss how we obtained these values from the TNG100 galaxies and we show that TNG100 successfully models the resolved main sequence. In addition, we introduce the $\Delta \Sigma_{\mathrm{SFR}}$ parameter to describe the offset of a simulated spaxel from the main sequence.

To obtain resolved maps of star formation, we first project the bound stellar particles for a given galaxy onto a 2D plane, where the plane is chosen such that the galaxy is viewed face-on. We use a cubic spline kernel, a standard choice for smoothed particle hydrodynamics, to map the particles onto the grid.  Here the grid spacing is $0.5 h^{-1}$~kpc, which corresponds to the gravitational softening length in TNG100 and roughly mimics the sampling of the MaNGA survey after spaxels are binned to increase S/N. \cite{Springel01a} define the kernel over $[0,h_{sml}]$ for an appropriate smoothing length, $h_{sml}$. 
We adopt the two dimensional spline kernel, with mass assignment function:

\begin{equation}
    W(r, h_{sml}) = \frac{40}{7\pi} \frac{1}{h_{sml}^2}
    \begin{cases}
        1 - 6q^2 + 6q^3     & \text{if } 0 \leq q \leq \frac{1}{2}, \\
        2(1-q)^3            & \text{if } \frac{1}{2} < q \leq 1, \\
        0                   & \text{if } q >1,
    \end{cases}
\end{equation}
where $r$ is the radial distance and $q \equiv r/h_{sml}$. For each particle, we define $h_{sml}$ to be the distance to the $N=64^{\rm th}$ nearest stellar particle, unless that distance is smaller than a grid cell, in which case we assign the minimum $h_{sml}$ such that every particle gets mapped into at least one cell. In their study of the optical morphologies of TNG100 galaxies, \cite{Rodriguez19} found that their results were insensitive to the choice of $N$ in the range $[4,64]$. We have tested $N=8$ and $N=16$ and have found no significant difference in the resultant main sequence fits. Here we choose $N=64$ for because it results in SFR maps with a higher filling factor but, as we discuss in Section \ref{sec:gv}, the filling factor for SFR maps with this choice of $N$ may still not be sufficient.
While the pixels on maps derived from the simulation are not true spaxels in the observational sense, we will refer to our grid bins as such.

The $z=0$ snapshot position of stellar particles will not be representative of the actual location where these stars formed, as they will have travelled a considerable distance since formation. However, this problem is not unique to simulations, and will affect observational tracers in a similar manner.

Once particles in each subhalo have been mapped, we have values of $\Sigma_*$ and $\Sigma_{\mathrm{SFR}}$ for each spaxel. For our analysis, we omit any spaxels with surface mass density $\Sigma_* < 10^6 \; \mathrm{M}_\odot \mathrm{kpc}^{-2}$. This is consistent with the analysis performed by \cite{Bluck2020b}, and corresponds roughly to the expected surface mass density of low-mass galaxies at $\sim2.5$ effective radii. For spaxels that have sufficient surface mass density, but $\Sigma_{\mathrm{SFR}}$ values that are either lower than the imposed sSFR limit, or are equal to zero, we adopt the same fixed minimum specific SFR as \cite{Bluck2020a}: sSFR = $10^{-12} \mathrm{yr}^{-1}$.

MaNGA spaxels are binned before analysis, with the goal of amplifying the S/N in each spaxel. Spaxels with high SFR will generally have higher S/N values.  In comparison to low-SFR spaxels, then, this procedure results in fewer high-SFR spaxels being binned together. Our simulated spaxels do not have S/N and are not binned in a similar manner. This results in a sample that is dominated by low-SFR spaxels, which makes it difficult to determine where star formation is occurring. To address this, we weight our spaxels by the exponent of their distance to the resolved main sequence line ($e^{\Sigma_{\mathrm{SFR}}}$) in order to emphasize the high-SFR sample.

In Figure \ref{fig:resMS} (left) we present the resolved main sequence ($\Sigma_* - \Sigma_{\rm{SFR}}$ relation) for our sample of spaxels. Hexagonal bins are used to group spaxels and are colored based on their logarithmic distance to the resolved main sequence line ($\Delta\Sigma_{\mathrm{SFR}}$). We limit the fit to the mode of star-forming spaxels, which are those that reside above the limit of $\log_{10} (\Sigma_{\mathrm{SFR}}/[M_{\odot}\mathrm{yr^{-1} kpc^{-2}}])>-3$. This is done to account for the numerous spaxels with SFR densities that were assigned the fixed lower limit, and which would otherwise overwhelm the mode of the distribution at most stellar mass densities. For this reason, and those described above, we draw contours to indicate the density of points, but weight the distribution from which the contours are drawn by $e^{\Delta \Sigma_{\mathrm{SFR}}}$. The right side of Figure \ref{fig:resMS} shows a bimodal distribution of star-forming and quenched spaxels where, again, weighting was performed in order to emphasize the contributions of star-forming spaxels to the distribution. As in Figure \ref{fig:glMS}, the distinct red peak is a result of the fixed minimum sSFR, rather than a genuine physical phenomenon. 
Were it possible to adequately resolve such low levels of star formation, one might expect that this distribution would extend out to very low values of $\Delta \Sigma_{\mathrm{SFR}}$, without a distinct peak. 

The resolved main sequence is identified with a least squares linear fit to the mode of the distribution of spaxels with $\log_{10} (\Sigma_{\mathrm{SFR}}/[M_{\odot}\mathrm{yr^{-1} kpc^{-2}}])>-3$ in bins of $\log_{10}(\Sigma_*/[M_{\odot} \mathrm{kpc^{-2}}])\sim0.08$. The fit is performed in the region $6.1 < \log_{10}(\Sigma_*/[M_{\odot} \mathrm{kpc^{-2}}]) < 7.8$, where the mode of the distribution is roughly linear. At surface mass densities $\log_{10}(\Sigma_*/[M_{\odot} \mathrm{kpc^{-2}}])>7.8$, the mode of the distribution plateaus at $\log_{10}(\Sigma_{\mathrm{SFR}}/[M_{\odot}\mathrm{yr^{-1} kpc^{-2}}]) \approx -1.2 $. 
We determine the resolved main sequence, with errors being the standard deviations of the fit, to be:

\begin{equation}
    \centering
    \log_{10} \Sigma_{\mathrm{SFR,MS}} = (0.91 \pm 0.03)\times \log_{10} \Sigma_* - (8.4 \pm 0.2).
    \label{eq:rSFMS}
\end{equation}
With this resolved main sequence, we assign a value of $\Delta\Sigma_{\mathrm{SFR}}$ to each spaxel as:
\begin{equation}
    \centering
    \Delta \Sigma_{\mathrm{SFR}} = \log_{10}(\Sigma_{\mathrm{SFR}}) - \log_{10}(\Sigma_{\mathrm{SFR,MS}}),
\end{equation}
which represents distance from the main sequence line, regardless of stellar mass, making it useful for comparing spaxels with a broad range of stellar mass densities. Table \ref{tab:rSFMS} summarizes slopes and normalizations that were obtained by previous observational and simulated studies. Providing for differences in assumptions and techniques, we find that the slope we have identified here is in good agreement with previously published results.
\begin{deluxetable*}{lcccccc}
    \floattable
    \tabletypesize{\scriptsize}
    \tablecolumns{7}
    \tablewidth{0pt}
    \tablecaption{Slopes and zero points for the $z=0$ resolved star formation main sequence (equation \ref{eq:rSFMS}) recovered from simulations and observations.
    \label{tab:rSFMS}}
    \tablehead{\colhead{Paper} & \colhead{Data Type} & \colhead{Data Source} & \colhead{Spaxel Resolution} & \colhead{Fitting Technique} & \colhead{Slope} & \colhead{Zero Point}} 
    \startdata
    this work & simulation  & TNG100 & $\sim 0.75$ kpc & ridge line & $0.91 \pm 0.03$  & $-8.4 \pm 0.2$ \\
    \cite{Trayford19} & simulation & EAGLE; $25^3$ Mpc$^3$ & $\sim 1.$ kpc  & OLS to median & $0.75$ & \\
    \cite{Trayford19} & simulation & EAGLE; $100^3$ Mpc$^3$  & $ \sim 1.$ kpc & OLS to median & $0.71$ & \\
    \cite{Hani20} & simulation & FIRE-2 & $0.1$ kpc & OLS & $0.52 \pm 0.0002$  & \\
    \cite{Hani20} & simulation & FIRE-2 & $0.5$ kpc & OLS &  $0.88 \pm 0.0014$ & \\
    \cite{Hani20} & simulation & FIRE-2 & $1.$ kpc & OLS &  $0.98 \pm 0.0027$ & \\
    \cite{Cano-Diaz16}  &  IFS & CALIFA DR2, $100\%$ of data  & $0.5$--$1.4$ kpc & OLS & $0.68 \pm 0.04$  & $-7.63 \pm 0.34$ \\
    \cite{Cano-Diaz16} & IFS  & CALIFA DR2, $80\%$ of data & $0.5$--$1.4$ kpc & OLS & $0.72 \pm 0.04$ & $-7.95 \pm 0.29$ \\
    \cite{Maragkoudakis16} & photometry  & SFRS & $\sim 1$ kpc & OLS &  $0.91 \pm 0.01$ & $-9.01 \pm 0.05$ \\
    \cite{Abdurro'uf17} & photometry & GALEX + SDSS & $1$--$2$ kpc &  ridge line & $1.00$  & $-9.58$ \\ 
    \cite{Liu_2018} & IFS  & MaNGA SDSS DR14 & $\sim 0.75$ kpc & OLS & $0.75$ &  $-8.19$ \\
    \cite{Erroz-Ferrer} & IFS  & MUSE Atlas of Disks (MAD) & $\sim 100$ pc & OLS & $0.714 \pm 0.064$ & $-7.56 \pm 0.2$ \\
    \cite{Bluck2020a} & IFS  & MaNGA SDSS DR15 & $\sim 0.75$ kpc & ridge line & $0.9 \pm 0.22$ & $-8.24 \pm 0.19$ 
    \enddata
    
    \tablecomments{All investigations in this table adopted a linear definition of the main sequence. OLS - Ordinary Least Squares}  

\end{deluxetable*}

\subsection{Radial Profiles} \label{sec:RP}
Here, we construct radial profiles of both $\Delta \Sigma_{\mathrm{SFR}}$ and luminosity-weighted age. In both cases we follow \cite{Bluck2020b} in normalizing radial distances by the half-light radius in the \textit{r}-band ($R_e$), as calculated by \cite{Genel2018} for the TNG100 simulation. Normalizing by this radius allows us to compare systems of varying size at the same relative position. Our bins have size $0.1R_e$, and we bin out to a maximum radial distance of $1.5R_e$. Altering the size of our bins does not substantially affect the resultant profiles or the conclusions we draw. 

Population-averaged profiles of $\Delta \Sigma_{\mathrm{SFR}}$ were constructed with the spaxels described in Section \ref{sec:lsf}. In a given radial bin, we take a weighted median of every spaxel in that bin for each galaxy population. The median is weighted by $e^{\Delta \Sigma_{\mathrm{SFR}}}$, for the reasons described in Section \ref{sec:lsf}.

 In addition, we compute population-averaged profiles of luminosity-weighted stellar age. Rather than smoothing particles onto a 2D grid, we directly project particles face-on into 2D radial bins, with no smoothing. For each bin, a luminosity-weighted age is then computed over all stellar particles belonging to the galaxy population that resides within that bin. In observations, a luminosity-weighted age is an indirect measure, requiring assumptions about the stellar population and star formation history. From simulations, we can compute the luminosity-weighted age directly since the formation time of each stellar particle is tracked. Here we compute the luminosity-weighted age using 
\begin{equation}
    \centering
    \langle x \rangle = \frac{\Sigma_k L_k x_k}{\Sigma_k L_k},
    \label{eq:age}
\end{equation}
where $x_k$ is $\log_{10}$ of the $k^{\rm th}$ particle's formation age, $L_k$ is the $r$-band luminosity of the $k^{\rm th}$ particle, and the summation is performed over all particles in each radial bin (see, e.g., \citealt{Gonzalez14}). 

There are a number of advantages to looking at both profiles. The $\rm \Delta \Sigma_{SFR}$ profiles have been constructed so as to be most comparable to observations (i.e., by smoothing particles onto a 2D grid and computing time-averaged SFRs). While we do not bin on signal to noise, we introduce a weighting factor to account for the emphasis that the MaNGA signal-to-noise binning technique places on high-SFR spaxels. The luminosity-weighted age makes no assumptions about the timescales over which observational tracers operate, as is the case for the $\rm \Delta \Sigma_{SFR}$ profiles. However, luminosity-weighted ages from observations are a less direct measurement. For these reasons, we will focus on comparing the general shapes and trends of these profiles, rather than the normalizations.  Following \cite{Bluck2020b} we compute a gradient for each profile, from the center out to a radial distance $1R_e$, beyond which the profiles often exhibit different behavior. We fit a linear function to the profiles out to $1 R_e$, take the slope to be the gradient ($\nabla_{1R_e}$), and the error on the slope to be one standard deviation from the fit.

\section{Results} \label{sec:results}
In this section, we present population-averaged radial profiles of luminosity-weighted ages (\age{}) and median offsets from the resolved main sequence ($\Delta\Sigma_{\mathrm{SFR}}$). By averaging profiles over different star-forming populations we can explore the connection between global SFR and SFR on local scales. In addition, since observations have found that the fraction of central galaxies that are quenched depends upon stellar mass (i.e., galaxies with high stellar masses are more likely to be quenched than galaxies with low stellar masses; e.g.,  \citealt{Peng10,Peng12}), we also explore \SFR{} and \age{} radial profiles as a function of stellar mass.  Finally, we compare the resolved SFR profiles of central and satellite galaxies in order to determine whether they undergo different quenching processes.

\begin{figure}[tp]
	\centering

	\includegraphics[width=\linewidth]{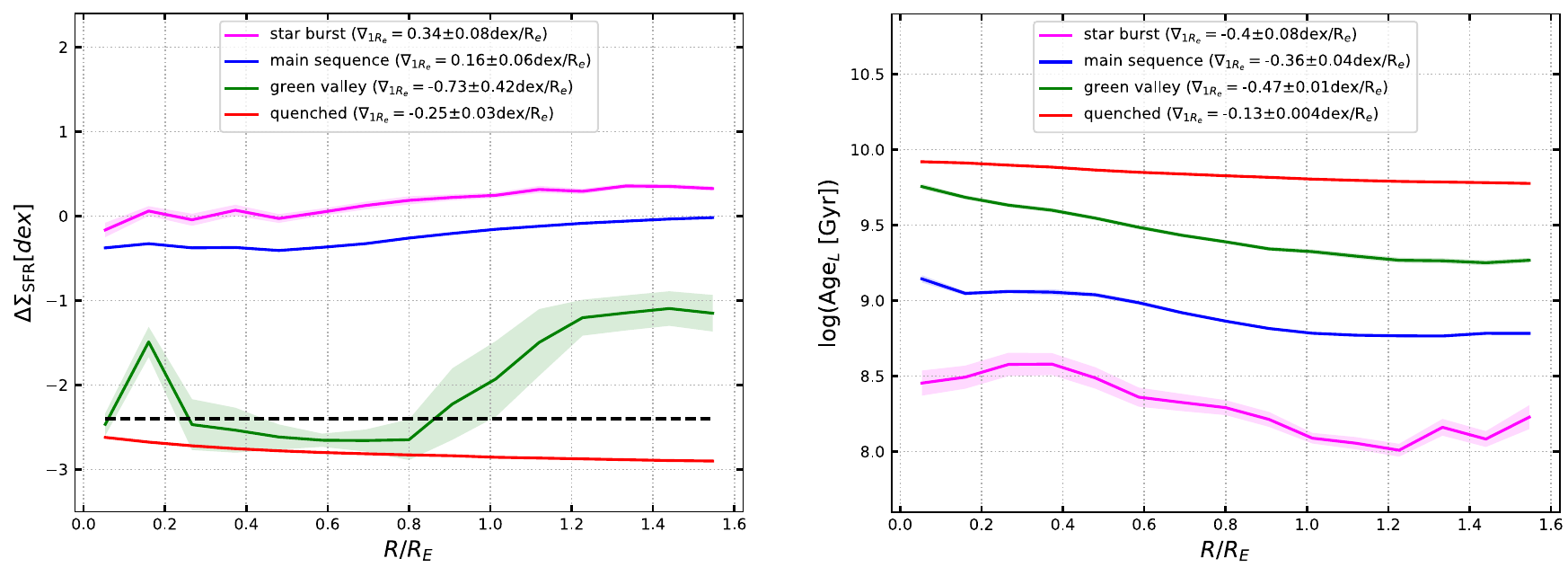}

    \caption{Population-averaged radial profiles of $\rm\Delta \Sigma_{SFR}$ (left) and \age{} (right). Profiles are shown separately for starburst (pink), main sequence (blue), green valley (green), and quenched (red) populations. Shaded regions indicate the error computed via $2000$ boostrap resamplings; where not visible, the errors are smaller than the width of the lines. The black dashed line on the left figure indicates the quenching threshold for spaxels (\SFR$=-2.4$), below which a spaxel is considered to be quenched. The legends include a profile gradient, the slope of a line fit to the profiles for $R \le 1R_e$. A positive gradient in \SFR{} is indicative of inside-out quenching, or, in the case of the starbursts, indicative of star bursting occurring in the outskirts. A negative gradient in \age{} is indicative of inside-out quenching, or star bursting in the outskirts.}
    \label{fig:allgal}
\end{figure}

In Figure \ref{fig:allgal}, we present radial profiles of $\rm \Delta \Sigma_{SFR}$ (left) and \age{} (right) for galaxies that are classified globally as starburst (purple), main sequence (blue), green valley (green), and quenched (red). Uncertainties were estimated via 2,000 bootstrap re-samplings of the data.  Uncertainties are represented by semi-transparent boundaries when they are larger than the widths of the lines and are omitted when they are comparable to or smaller than the widths of the lines. Legends in Figure \ref{fig:allgal} list the gradients for each profile, computed from the centers of the galaxies out to radial distances corresponding to $1 R_e$. The minimum of the $\Delta\Sigma_{\mathrm{SFR}}$ distribution (Figure \ref{fig:resMS}, right inset) defines the threshold below which we consider a spaxel to be definitively quenched and this is indicated by the black dashed line in Figure~\ref{fig:allgal}. 

\begin{figure}
    \centering
    \includegraphics[width=\linewidth]{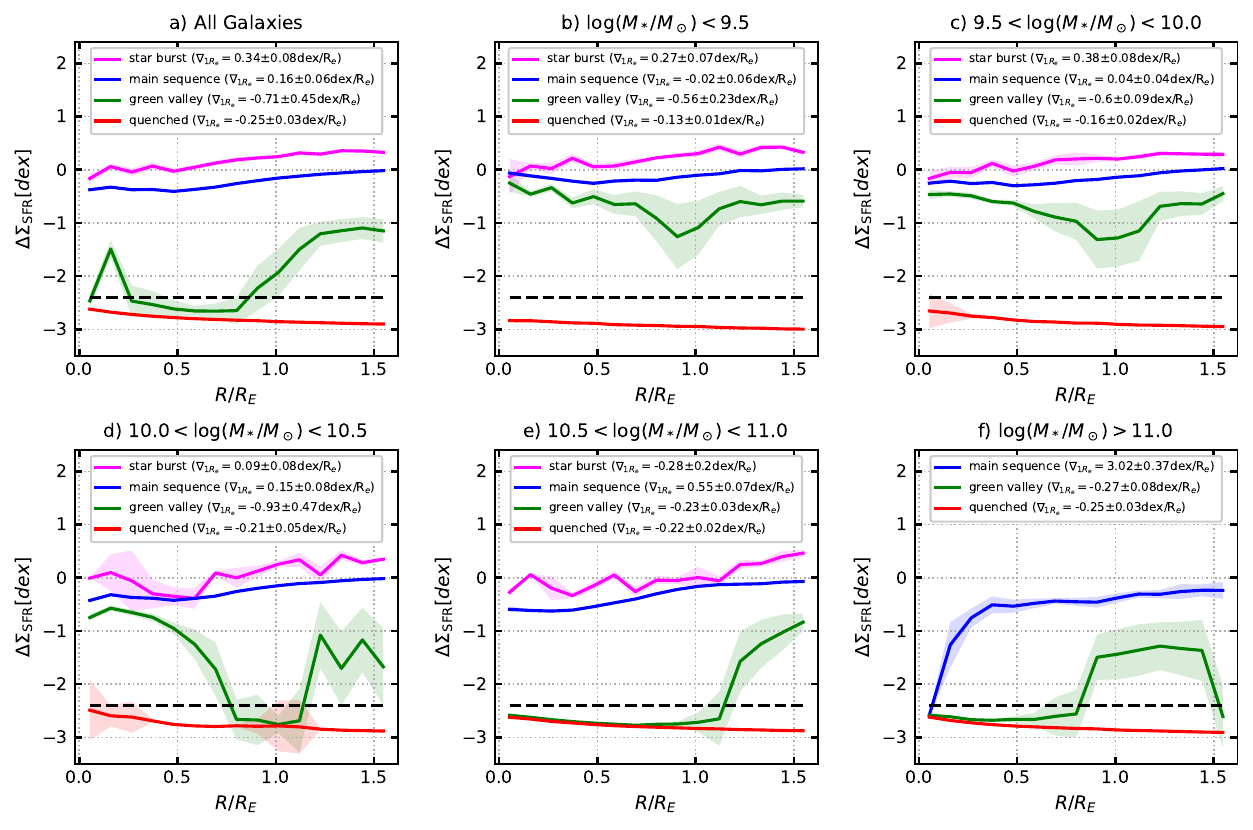}
    
    \caption{Dependence of \SFR{} radial profile on stellar mass. Formatting is identical to Figure \ref{fig:allgal}, Panel a) includes all galaxies and is a reproduction of Figure \ref{fig:allgal} (left). With the exception of the starburst galaxies in panels d) and e), which each contain only $3$ objects, at least $20$ galaxies contribute to each individual profile. Note: there are no starburst galaxies in the mass range of panel f. Errors are computed from 2000 bootstrap resamplings. Error bounds that are wider than the linewidth are indicated by shaded regions and are omitted from the figure when they are comparable to or smaller than the linewidth.}
    \label{fig:bymass}

\end{figure}

\begin{figure}
    \centering
    \includegraphics[width=\linewidth]{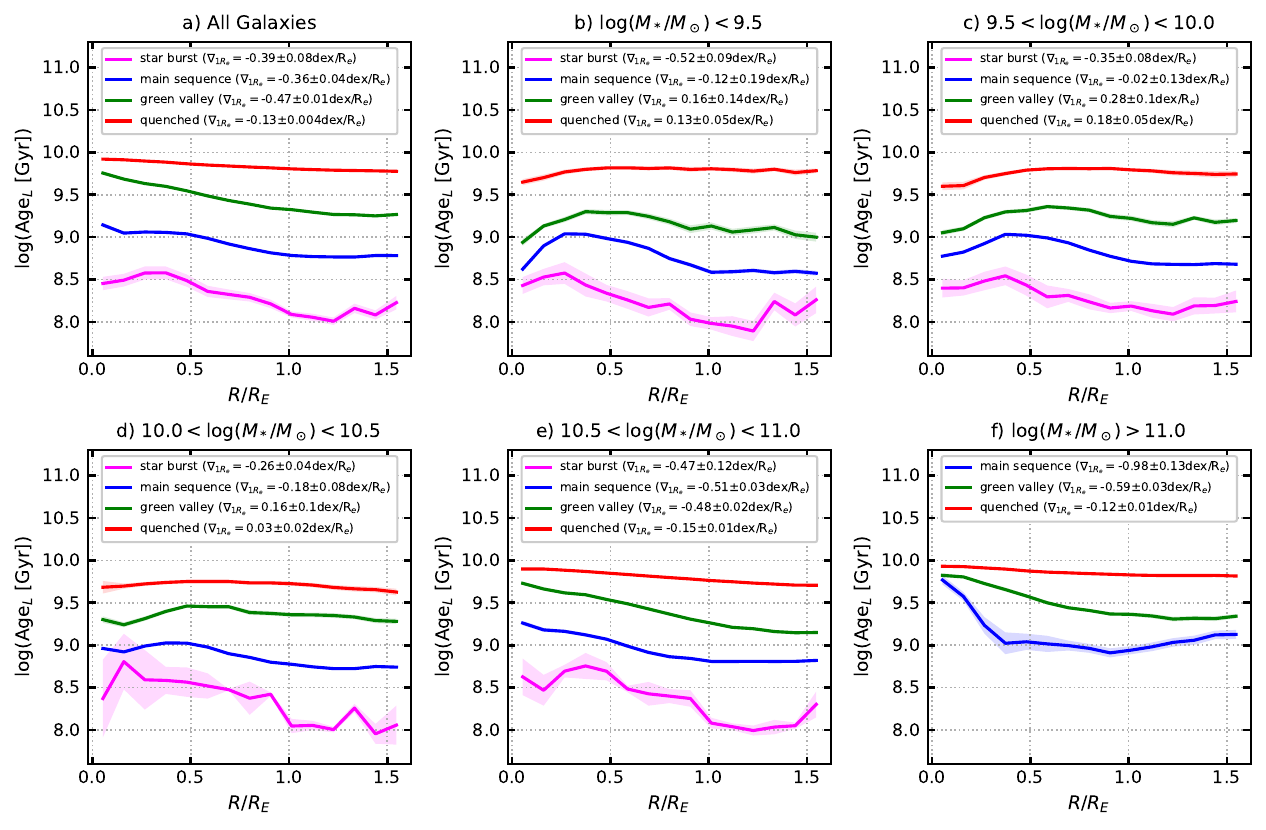}
    \caption{Dependence of \age{} radial profile on stellar mass. Formatting is identical to Figure \ref{fig:allgal}. Panel a) includes all galaxies and is a reproduction of Figure \ref{fig:allgal} (right). With the exception of the starburst profiles in panels d) and e), at least $20$ galaxies contribute to each individual profile. In all cases, however, the value of \age{} of each radial bin is computed using at least 1000 particles. Note: there are no starburst galaxies in the highest mass bin. Errors are computed from 2000 bootstrap resamplings. Error bounds that are wider than the linewidth are indicated by shaded regions and are omitted from the figure when they are comparable to or smaller than the linewidth.}
    \label{fig:age_bymass}
\end{figure}

Figures \ref{fig:bymass} and \ref{fig:age_bymass} show, respectively, profiles of \SFR{} and \age{} for the galaxies from Figure \ref{fig:allgal}, subdivided by stellar mass. 
Gradients for each of the profiles in Figures \ref{fig:allgal}, \ref{fig:bymass}, and \ref{fig:age_bymass} are listed in the legends. With the exception of the starburst population (which we will discuss below), a positive gradient in $\rm \Delta \Sigma_{SFR}$ indicates inside-out quenching, with the outskirts of the galaxies showing a greater amount of star formation relative to the resolved main sequence than is shown by their centers. Inside-out quenching is also indicated by a negative \age{} gradient, which reflects older stellar particles in the center of a galaxy, with younger stellar particles residing in its outskirts. That is, the gradients in the $\rm \Delta \Sigma_{SFR}$ and $\rm Age_L$ profiles should have opposite signs within each population, with positive \SFR{} gradients corresponding to negative \age{} gradients. 
For particularly noisy \SFR{} profiles, the \SFR{} gradient may be less trustworthy than the gradient derived from the analogous \age{} profile.

We discuss results for each population of galaxies in the following subsections. 

\subsection{Quenched Galaxies} \label{sec:quenched}
As expected, profiles of the quenched population consistently have the lowest values of \SFR{} and the oldest luminosity-weighted ages. In Figures \ref{fig:allgal} and \ref{fig:bymass}, the local \SFR{} values are below the quenching threshold, with a slightly negative gradient. Due to limited resolution, the \SFR{} profile of quenched galaxies is dominated by spaxels with a $\rm \Delta \Sigma_{SFR}$ value that corresponds to our imposed upper limit of sSFR$ = 10^{-12} \mathrm{yr}^{-1}$. Therefore, the quenched \SFR{} profiles should be seen as placeholders for arbitrarily low values of star formation and do not provide substantive information about the quenching histories of these objects.

In contrast to the \SFR{} profiles, the \age{} profiles of quenched galaxies are not affected by the sSFR upper limit and they do provide information about the quenching history.  When we consider all quenched galaxies (Figure \ref{fig:allgal}), the \age{} profile has a negative gradient, consistent with inside-out quenching. While accretion of younger stars in the outskirts of these galaxies would also explain the negative \age{} gradient, fossil record analysis indicates that inside-out quenching is the likely driver of similar trends in quiescent MaNGA galaxies \citep{Avila-Reese}.  

From Figure \ref{fig:age_bymass}, it is clear that the \age{} profiles for quenched galaxies depend on the stellar masses of the galaxies. That is, low-mass quenched galaxies with $\log_{10} M_*/M_\odot <10.0$ (panels b and c) have positive \age{} gradients, quenched galaxies with $10.0 < \log_{10} M_*/M_\odot < 10.5$, have flat \age{} profiles on average, and high-mass quenched galaxies ($\log_{10} M_*/M_\odot >10.5$) have negative \age{} gradients. This result is consistent with the inside-out quenching scenario in which quenching is driven via an amount of feedback that scales with galaxy stellar mass. 

\subsection{Green Valley Galaxies} \label{sec:gv}
Profiles of star formation in green valley galaxies are of particular interest because these objects are in a transition stage between being main sequence galaxies that are actively forming stars and quenched systems that are no longer forming stars. 
From the left panel of Figure \ref{fig:allgal}, the \SFR{} profile for green valley galaxies exhibits some distinct features.
At intermediate radii ($R/R_e \approx 0.3-0.8$), the \SFR{} for these objects is at its minimum, exhibiting low levels of star formation that only slightly exceed the \SFR{} of the quenched galaxies, due to a low filling factor of star-forming spaxels at these radii.   The trend in \SFR{} profile shape is not reflected in the corresponding \age{} profile for these objects (Figure \ref{fig:allgal}, right panel). It should also be noted that the green valley phase is relatively short-lived, resulting in more modest changes to the \age{} profile than to the \SFR{} profile. 
For radii greater than $0.8 R_e$, the \SFR{} and \age{} profiles in Figure \ref{fig:allgal} both indicate that the star formation rates are greater and the ages are younger than they are at smaller radii, as would be expected for inside-out quenching. 
The shape of the \SFR{} profile for green valley galaxies is clearly not linear, and this is reflected in the large error for the gradient ($-0.84 \pm 0.48$). However, the gradient of the green valley \age{} profile is indicative of inside-out quenching, with \grad(\age) $= -0.47 \pm 0.01$. 

Of all the \SFR{} profiles in the left panel of Figure \ref{fig:allgal}, the profile for the green valley population has the largest error bounds, indicating a greater degree of variation amongst the profiles for individual green valley galaxies. Indeed, it is clear from Figures \ref{fig:bymass} and \ref{fig:age_bymass} that the \SFR{} and \age{} profiles for green valley galaxies are strongly dependent on stellar mass. Profiles of green valley galaxies with $\log_{10} M_*/M_\odot <10.5$ (Figure \ref{fig:bymass}b,c) show a decrease in \SFR{} at radii $\sim 1 R_e$ that is not accompanied by a corresponding increase in luminosity-weighted age in Figure \ref{fig:age_bymass}b,c. Thus, the decline in \SFR{} at these radii is likely not significant. 
The \SFR{} profiles at the centers of green valley galaxies with $\log_{10} M_*/M_\odot >10.5$ (Figure \ref{fig:bymass}e,f) are below the quenching threshold, but star formation is still detected in the outskirts of these galaxies. 
The gradients of these profiles do not reflect this, as they are only fit out to $1 R_e$.

The \age{} profiles of green valley galaxies are not affected by issues with filling factors since they are obtained by binning stellar particles directly. For green valley galaxies with $\log_{10} M_*<10.5$, the \age{} profiles (Figure \ref{fig:age_bymass}b,c,d) rise and fall, with the oldest luminosity-weighted ages occurring at $\sim 0.5 R_e$. Thus, the youngest, brightest stars are primarily located at the centers and the outskirts of these galaxies. Green valley galaxies with $\log_{10} M_*/M_\odot >10.5$ (Figure \ref{fig:age_bymass}f,e) have decreasing \age{} profiles, with the oldest luminosity-weighted ages at the centers. The gradient of the \age{} profile for galaxies with the highest stellar masses (Figure \ref{fig:age_bymass}f) is significantly steeper than than it is for galaxies in the next highest mass bin (Figure \ref{fig:age_bymass}e).  This is consistent with inside-out quenching, in which quenching is driven by central feedback that increases in strength with increasing galaxy mass.

Lastly, from Figure \ref{fig:age_bymass} we note that, in general, the inner regions of more massive green valley galaxies tend to have older luminosity-weighted ages than do the inner regions of less massive green valley galaxies. This indicates that more massive green valley galaxies have centers that quenched earlier than their less massive counterparts, possibly due to the massive galaxies being older on average or the quenching process having begun earlier in these galaxies.

\subsection{Main Sequence Galaxies} \label{sec:ms}
From Figure \ref{fig:allgal}, the main sequence population has profiles with a positive gradient in \SFR{} ($0.16 \pm 0.06$) and a negative gradient in \age{} ($-0.36 \pm 0.04$), consistent with inside-out quenching on average.
However, it is clear from Figure \ref{fig:bymass}, that the \SFR{} profiles of the main sequence galaxies have a strong dependence on total stellar mass. Main sequence galaxies with $\log_{10} M_*/M_\odot <10$ (panels b and c) have relatively flat gradients measured out to $1.0 R_e$.
At higher masses, the steepness of the profile gradients increases with increasing stellar mass; i.e., high mass main sequence galaxies exhibit more central suppression of star formation than their low mass counterparts, consistent with inside-out quenching for galaxies with $\log_{10} M_\ast/M_\odot > 10$. 

The dependence of the \age{} profiles of main sequence galaxies with $\log_{10} M_*/M_\odot >10$ (Figure \ref{fig:age_bymass}, panels d, e, and f) on stellar mass is opposite to that of the \SFR{} profiles, with negative gradients for \age{} that become steeper with increasing stellar mass. 
The shapes of \age{} profiles for the $\log_{10} M_*/M_\odot <10$ main sequence galaxies are not well described by a linear gradient, as they exhibit younger luminosity-weighted ages at the very center, with the average age peaking at $\sim 0.4 R_e$. Changing bin size does not significantly alter these results.


\subsection{Starburst Galaxies} \label{sec:sb}
From Figure \ref{fig:allgal}, the \SFR{} and \age{} profiles for the starburst population have shapes and gradients that are similar to those of the main sequence population. However, gradients of the radial profiles of starburst galaxies should be interpreted differently from those of other populations. This is because starburst galaxies are not actively quenching; rather, they are undergoing bursts of star formation. Thus, higher values of $\rm \Delta \Sigma_{SFR}$ and lower values of \age{} indicate where the star bursting is occurring locally within the galaxies, instead of where quenching has yet to occur. In Figure \ref{fig:allgal}, higher \SFR{} and lower \age{} values occur at the outskirts, indicating that the star bursting is occurring primarily in the outer regions of the TNG100 galaxies. 

With only 58 galaxies in total, the starburst population is considerably smaller than the other populations. However, each starburst galaxy has at least $300$ spaxels contributing to the \SFR{} profile and each radial bin in the \age{} profiles has at least 1,000 stellar particles, with most bins having tens of thousands.

In Figures \ref{fig:bymass} and \ref{fig:age_bymass}, there are $31$ starburst galaxies with $\log_{10} M_*/M_\odot <9.5$ (panel b), $21$ starburst galaxies with $9.5< \log_{10} M_*/M_\odot < 10.0$ (panel c), and $3$ starburst galaxies each with the $10.0 < \log_{10} M_*/M_\odot <10.5$ (panel d) and $10.5 < \log_{10} M_*/M_\odot < 11.0$ (panel e). There are no TNG100 starburst galaxies with $\log_{10} M_*/M_\odot >11.0$. 
From Figures \ref{fig:bymass} and \ref{fig:age_bymass}, the shapes and gradients of the \SFR{} and \age{} profiles show no clear trend with stellar mass.  What is clear, however, is that the starburst \SFR{} profiles increase with radius and their corresponding \age{} profiles decrease with radius. Thus, unlike what is seen for observed starburst galaxies \citep[e.g.,][]{Bluck2020b,Ellison18}, the star bursting in the TNG100 galaxies is not centrally concentrated. This is likely due to limitations in modeling star bursting galaxies within cosmological-volume simulations \citep[see][]{Sparre16}.

\subsection{Insights from Different Galaxy Populations}
The profiles of quenched, green valley, and main sequence galaxies in Figures \ref{fig:allgal}, \ref{fig:bymass}, and \ref{fig:age_bymass} are consistent with the inside-out quenching scenario. That is, massive ($\log_{10} M_*/M_\odot >10.5$) galaxies that are undergoing quenching show suppression of star formation in their inner regions relative to their outskirts. Massive green valley galaxies also show suppression of star formation throughout, as compared to galaxies that remain on the $z=0$ main sequence. Indeed, there is excellent agreement between the global classification of galaxies and their local \SFR{} and \age{} profiles. That is, starburst galaxies have higher \SFR{} and younger \age{} than main sequence galaxies, main sequence galaxies have higher \SFR{} and younger \age{} than green valley galaxies, and quenched galaxies have the lowest \SFR{} and oldest \age.

Since there are regions where the filling factor for \SFR{} appears to be insufficient in green valley galaxies, we will refrain from showing further profiles of this parameter. We have, however, confirmed that the insights we obtain from the \age{} profiles are also reflected in the central regions ($\lesssim 0.5 R_e$) of the corresponding \SFR{} profiles, where the filling factor does appear to be sufficient. Thus, while changes in how the maps of \SFR{} are generated may provide more robust profiles at larger radii, such changes would not qualitatively change our main results.

\subsection{Central and Satellite Galaxies} \label{sec:centsat}

\begin{figure}
    \centering
    \includegraphics[width=0.7\linewidth]{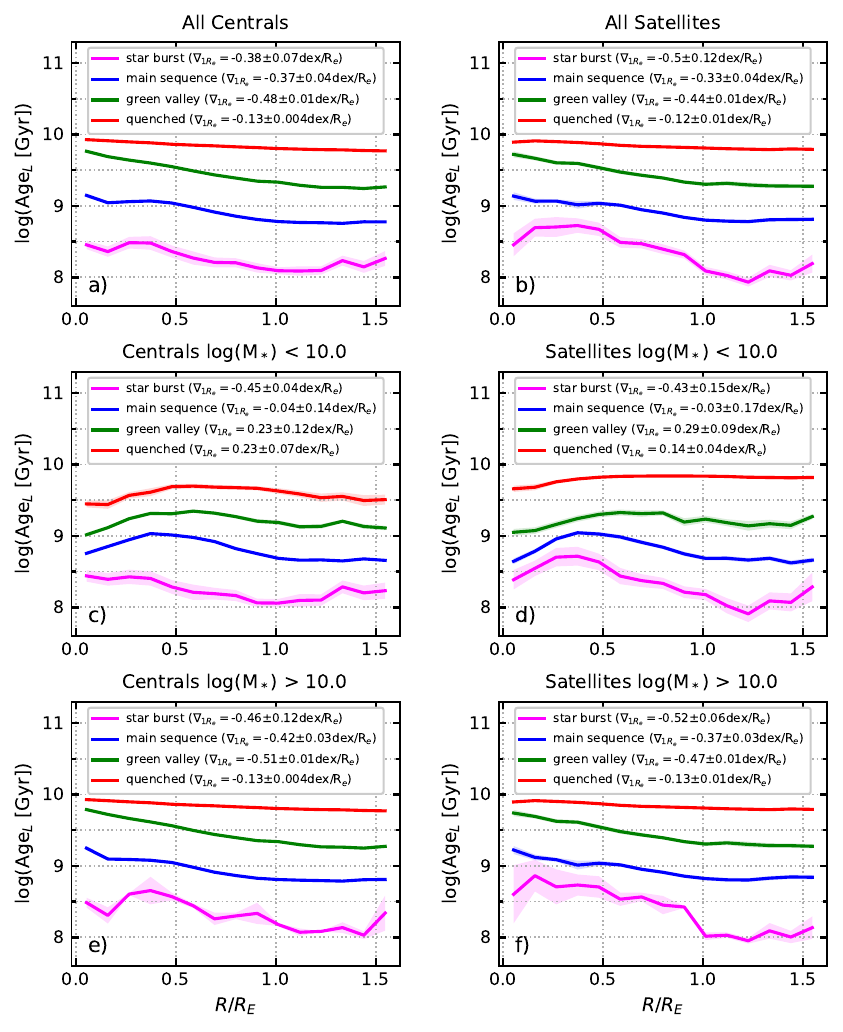}
    \caption{Radial profiles of \age{}, with galaxies separated into centrals (left) and satellites (right). Formatting is identical to Figure \ref{fig:allgal} (right). Population-averaged profiles are shown in the top row for all centrals (panel a) and all satellites (panel b), and subsequent rows show profiles for the low-mass ($\log_{10} M_*/M_\odot <10$, middle) and high-mass ($\log_{10} M_*/M_\odot > 10$, bottom) subsets of the central and satellite populations. All populations except for the high- and low-mass starbursts have at least 20 contributing galaxies. There are only $2$ ($36$) low- (high-) mass starburst centrals and only $16$ ($4$) low- (high-) mass starburst satellites. All bins are computed with at least $1000$ stellar particles regardless of how many galaxies contribute. Errors, represented by the shaded regions, are computed via 2000 bootstrap resamplings, and are smaller than the width of the lines when not visible.}
    \label{fig:centsat}
\end{figure}

Different quenching mechanisms are thought to operate on central and satellite galaxies \citep[e.g.,][]{Bluck2020b}. In Figure \ref{fig:centsat}, we present the radial profiles of \age{} for all centrals (panel a) and satellites (panel b). The bottom two rows further divide the sample into low- ($M_* < 10^{10} M_\odot$)and high-mass ($M_* > 10^{10} M_\odot$) systems.  %

In the top panels of Figure \ref{fig:centsat}, the profiles of \age{} have negative slopes for all types of central and satellite galaxies.  The gradients of the \age{} profiles for main sequence central and satellite galaxies agree well, as do the gradients of the \age{} profiles for quenched central and satellite galaxies. The gradient of the \age{} profile for green valley central galaxies is, however, somewhat steeper (\grad $=-0.48\pm0.01$) than it is for green valley satellite galaxies (\grad $=-0.44\pm0.01$). In the case of star bursting galaxies, the gradient of the \age{} profile is steeper for satellite galaxies (\grad $=-0.5\pm0.12$) than it is for central galaxies (\grad $=-0.38\pm0.07$).  

As Figure \ref{fig:age_bymass} demonstrates that radial profiles of star formation vary with mass, we further divide our sample of central and satellite galaxies by mass. 
The \age{} profiles for these sub-samples are presented in the bottom two rows of Figure \ref{fig:centsat}. There is a clear difference for both central (left) and satellite (right) populations between the profiles of low-mass galaxies (middle panels) and high-mass galaxies (bottom panels). The bottom panels of Figure \ref{fig:centsat} demonstrate what was already seen in Figure \ref{fig:age_bymass}: sufficiently massive green valley and main sequence galaxies have negative \age{} gradients, indicative of inside-out quenching. There are very few starburst galaxies in this high mass bin:  only two (four) central (satellite) star burst galaxies. We therefore refrain from interpreting these in the manner of other population-averages, but show them for completeness. 

Panels c) and d) of Figure \ref{fig:centsat} show \age{} profiles for low-mass centrals (left) and satellites (right). In both low-mass centrals and satellites, the central value of \age{} is smaller than it is for the surrounding regions, in contrast to the smoothly declining age profiles of the high-mass centrals and satellites. The centers of the low-mass green valley centrals and satellites are actually younger than any other region within the galaxies. This indicates that, at least until very recently, the centers of these populations were still actively forming stars. Thus, mass quenching does not appear to be the dominant process that quenches low-mass galaxies, whether they are centrals or satellites.

The normalization of \age{} across populations largely depends on star-forming type and galaxy stellar mass, rather than whether a galaxy is a central or satellite. That is, for the high-mass population (bottom panels), the median ages of quenched, green valley, main sequence, and even starburst populations are roughly similar for centrals and satellites at the same relative radii. This is also the case with the low-mass population (middle panels), with the notable exception of the quenched population. The radial bins of low-mass, quenched satellites have luminosity-weighted ages ranging from $\sim9.6$--$9.8$~Gyr, while low-mass, quenched centrals have binned ages ranging from $\sim9.4$--$9.7$~Gyr. Of all quenched centrals, only $4\%$ have stellar masses $<10^{10} M_\odot$; for quenched satellites, $26\%$ are low-mass. Of all low-mass centrals, only $1.9\%$ are quenched, while $29\%$ of all low-mass satellites are quenched.

Although we do not show profiles of \SFR{} for central and satellite galaxies here, we note that the \SFR{} profiles for low-mass central and satellite galaxies have similar shapes. For high-mass green valley satellites, there does appear to be higher \SFR{} at $\sim 0.5-1.0 R_e$ than for the corresponding central population, however the significance of this is unclear due to the poor filling factor at these radii.

Figure \ref{fig:centsat} shows that how a galaxy quenches is largely dependent its stellar mass; without sufficient mass to self-regulate star-formation through intrinsic quenching mechanisms, a galaxy will continue to form stars at its center. The radial profiles of main sequence and green valley low-mass galaxies appear remarkably similar, indicating that the quenching process may be similar for both centrals and satellites.
The older ages of low-mass, quenched satellites indicate that this population quenched earlier on average than the low-mass, quenched centrals.


In summary, our results illustrate how galaxies quench in the TNG100 simulation, and that the major factor in predicting how a galaxy will quench is its stellar mass. Massive ($\log_{10} M_*/M_\odot \gtrsim 10.5$) galaxies will quench from the inside-out, regardless of whether they are centrals or satellites. Low-mass galaxies ($\log_{10} M_*/M_\odot < 10.0$) do not undergo simple inside-out quenching, revealing relatively young ages at their centers. 
Further investigation of processes other than AGN feedback that may drive quenching, such as morphological or environmental factors, is of interest for understanding quenching in low-mass central and satellite galaxies.
 

\section{Comparison with Previous Studies} \label{sec:discussion}

In this section we compare our global and resolved star-forming main sequences, as well as the radial profiles for \age{} and \SFR{}, to previous results. It is, however, important to note that there are a number of challenges with direct comparisons of these types of results. First, the star formation tracers that are used in observational studies, the assumptions that are made in deriving the SFRs from the tracers, and different sample selection criteria can affect the normalization (i.e., the ``zero point'') of the main sequence. In addition, we adopted a definition for the main sequence that is linear in logarithmic space. Some studies \citep[e.g.,][]{Popesso18} have found that the main sequence deviates from a linear relation for galaxies with high stellar masses, for which it flattens out. Consequently, the main sequence is sometimes
fitted with a quadratic equation in these studies. Below, we restrict our comparisons to studies in which the main sequence was defined to be a linear relation in logarithmic space.

\subsection{Comparison of the Global Main Sequence} \label{sec:compGMS}

Table \ref{tab:SFMS} lists our results for the TNG100 global main sequence, along with the results of three observational studies \citep{Speagle_2014,Renzini15,Cano-Diaz16} and one study based on the largest-volume simulation in the TNG suite, TNG300 \citep{erratum}. From Table \ref{tab:SFMS},
the slope of the global main sequence that we find for TNG100 galaxies agrees well with the results of \cite{Speagle_2014} and \cite{Cano-Diaz16}, and is within $\sim 2\sigma$ of the results of \cite{Renzini15}. 
The zero point we obtain from TNG100 galaxies is in good agreement with the zero points obtained by \cite{Cano-Diaz16} and \cite{Renzini15}, but disagrees with that found by \cite{Speagle_2014} at the $\sim 3.5\sigma$ level. Given the large number of variables that can affect the zero point values, and evidence that slopes derived from ordinary least squares (OLS) can be biased by differences in the criteria used to select the galaxies \citep[see][]{Renzini15}, we find that, overall, the global main sequence of TNG100 galaxies is in good agreement with the global main sequence of observed galaxies.

From Table \ref{tab:SFMS}, our result for the global main sequence of TNG100 galaxies is in good agreement with that of \cite{erratum}, who investigated the global main sequence of TNG300 using time-averaged star formation rates. The agreement between our result and that of \cite{erratum} occurs despite some substantial differences in the investigations, including different timescales over which the SFRs were averaged (200~Myrs in \citealt{erratum} vs.\ 20 and 100~Myrs here), the use of somewhat different apertures within the galaxies when computing SFRs, somewhat different stellar masses ($> 10^{10.2} M_\odot$ in \citealt{erratum} vs.\ $>10^{10.4} M_\odot$ here), and different analysis techniques (performing an OLS fit to the global main sequence of galaxies that were pre-selected to be star-forming in \cite{erratum} vs.\ the ridge line definition adopted here).  That our global main sequence for TNG100 galaxies and that of \cite{erratum} agree suggests that the results are robust to modest differences in the approaches that are used to obtain the global main sequence.

\subsection{Comparison of the Resolved Main Sequence} \label{sec:CompRMS}

Table \ref{tab:rSFMS} shows our results for the resolved star formation main sequence of TNG100 galaxies, along with previous results from observed and simulated galaxies.  The range of slopes and zero points in Table \ref{tab:rSFMS} are reflective of the sensitivity the rSFMS to the SFR timescales that are used, the analysis technique, and other factors. In particular, galaxy selection criteria have an especially pronounced affect when the rSFMS is fit with ordinary least squares (OLS; see \citealt{Abdurro'uf17}).  In addition, the spatial resolution of individual spaxels affects the slope of the rSFMS \citep{Hani20}. Due to the clumpy nature of star-forming regions, the slope of the rSFMS becomes shallower as the spatial resolution is increased (i.e., as spatial resolution increases, $\Sigma_*$ remains relatively unchanged, but isolated star forming regions are spread over more pixels, effectively decreasing $\rm \Sigma_{SFR}$ in individual pixels; see \citealt{Hani20}).  

Because of the number of issues that can affect the rSFMS, the observational result that is most straightforward to compare directly to our result is that of \cite{Bluck2020a} (i.e., similar spatial resolution, and use of the ridge line to identify the rSFMS).  From Table \ref{tab:rSFMS}, our result for the rSFMS for TNG100 galaxies is in good agreement with the results of \cite{Bluck2020a} for observed galaxies.

None of the previous results for the rSFMS from simulations match the resolution of our work, and none used the ridge line to identify the rSFMS (see Table \ref{tab:rSFMS}).  However, it is interesting to note that the slope of the rSFMS we find for TNG100 galaxies (with spaxel resolution of $\sim 0.75$~kpc) falls between the slopes obtained for FIRE-2 galaxies using spaxel resolutions of 0.5~kpc and 1.0~kpc.  While there are no previous results from simulations that can be compared to ours as directly as the observational results of \cite{Bluck2020a}, it does appear that our result for the rSFMS of TNG100 galaxies is in reasonable agreement with that of the FIRE-2 galaxies.

\subsection{Comparison of Radial Profiles}\label{sec:compRP}

Here we compare our results for the radial profiles of \SFR{} and \age{} for TNG100 galaxies to results from the observational investigation that adopted definitions and analysis techniques that are the most similar to those that we adopted: \cite{Bluck2020b}.  Given the resolution issues that affect the accurate modeling of starburst galaxies within large-volume cosmological simulations, here we focus only on comparisons of results for main sequence, green valley, and quenched galaxies.  Results for the gradients of \SFR{} and \age{} profiles for our TNG100 galaxies and for observed galaxies in \cite{Bluck2020b} are shown in Fig.~\ref{fig:grad_all}.  Results for the gradient of the \age{} profile are shown in the right panel, while the left panel shows the {\it negative} of the gradient of the \SFR{} profile.  Dotted lines in both panels discriminate outside-in quenching from inside-out quenching.  In comparing these results, we note that we obtained gradients of the profiles by obtaining best-fit lines, while \cite{Bluck2020b} obtained gradients by simply comparing the values of a given parameter (\SFR{} or \age{}) at $R/R_e = 1$ and $R/R_e = 0$.

\subsubsection{Main Sequence Galaxies}


\begin{figure}
    \centering
    \includegraphics[width=\linewidth]{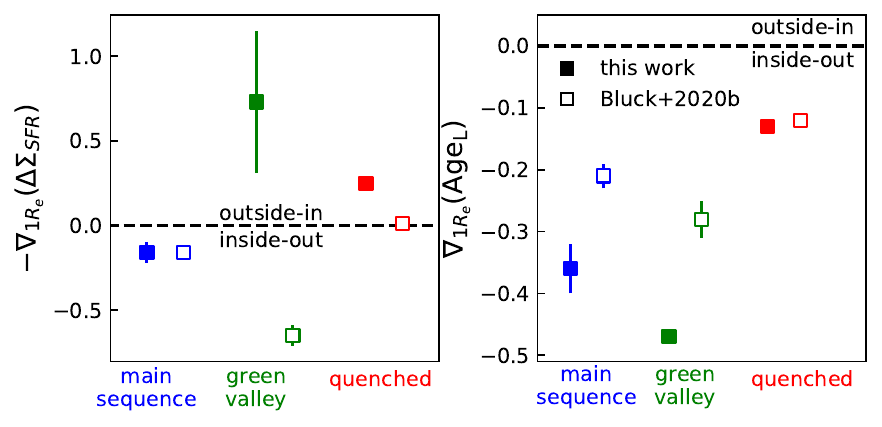}
    \caption{Profile gradients from the legend of Figure \ref{fig:allgal} in this work (closed squares) and from Figure 3 in \cite{Bluck2020b} (open squares). The left panel shows the \textit{negative} of the \SFR{} gradient, while the right panel shows the \age{} gradient. The dashed black lines are located at \grad(\SFR)$=0$ and \grad(\age)$=0$, where a profile would be flat.}
    \label{fig:grad_all}
\end{figure}

From Fig.~\ref{fig:grad_all}, there is good agreement between the quenching of main sequence TNG galaxies and observed galaxies (i.e., both are consistent with inside-out quenching).  The values of the gradients of the \SFR{} profiles for main sequence galaxies are also in good agreement. In addition, when the profiles of main sequence galaxies are computed separately for galaxies in different stellar mass bins, the most massive galaxy bin has the steepest \SFR{} gradient for both the TNG100 galaxies (see Fig.~\ref{fig:bymass}) and the galaxies in \cite{Bluck2020b}. 

While the gradients of the \SFR{} profiles of TNG100 main sequence galaxies and observed main sequence galaxies agree well, the gradient of the \age{} profile for TNG100 main sequence galaxies is somewhat steeper than that of observed galaxies. This may be explained by limitations in determining stellar age from observed spectra (i.e., the assumption of a single stellar population) or limitations in modeling low levels of star formation in simulations.


\subsubsection{Green Valley Galaxies}

The unreliability of the \SFR{} profiles of green valley TNG100 galaxies has already been discussed in Section \ref{sec:gv}.
From profiles of \age{} for this population,
we conclude that our results for TNG100 green valley galaxies are in qualitative agreement with \cite{Bluck2020b} (i.e., inside-out quenching). As in the case of main sequence galaxies, the gradient of the \age{} profile for TNG100 green valley galaxies is significantly steeper than that of the green valley galaxies in \cite{Bluck2020b}.

We find that the \age{} profiles of TNG100 green valley galaxies as a function of stellar mass agree with the general trends for observed green valley galaxies in \cite{Bluck2020b}: quenching proceeds inside-out in massive ($\log_{10}M_*/M_\odot > 10.5$) green valley galaxies, while centrally-driven quenching does not appear to be the dominant process in low-mass green valley galaxies.

\subsubsection{Quenched Galaxies}

The profiles of \age{} show that both TNG100 quenched galaxies and quenched galaxies in \cite{Bluck2020b} have slightly older stellar populations in their centers than in their outskirts.  The gradients of the two \age{} profiles agree within $1\sigma$, which is a considerably better agreement than we find for the gradients of the \age{} profiles for the main sequence and green valley galaxies.  This may indicate that a substantial contributor to the disagreement between the gradients of the \age{} profiles for observed vs. TNG100 main sequence and green valley galaxies is the assumption of a single stellar population within a given spaxel (i.e., quenched galaxies are likely to be modelled better by single stellar populations than are star-forming galaxies).

\subsubsection{Quenching in central and satellite galaxies}
\begin{figure}
    \centering
    \includegraphics[width=\linewidth]{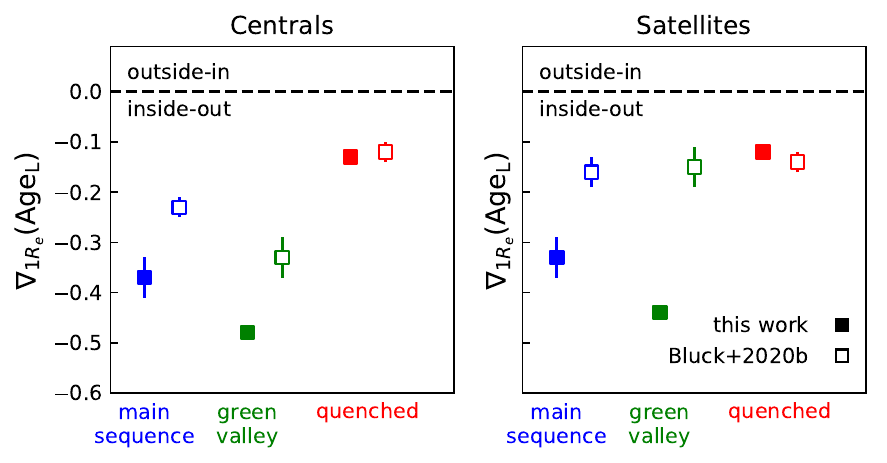}
    \caption{Profile gradients from the legend of Figure \ref{fig:centsat} in this work (closed squares) and from Figure 3 in \cite{Bluck2020b} (open squares). Gradients are plotted for centrals (left) and satellites (right), both panels share a y-axis. The dashed black lines are located at \grad\age$=0$, where a profile would be flat.}
    \label{fig:grad_centsat}
\end{figure}

Fig.~\ref{fig:grad_centsat} summarizes the gradients of the \age{} profiles for central galaxies (left) and satellite galaxies (right) in TNG100 and \cite{Bluck2020b}.  From this figure, there is good agreement between the observed and simulated galaxies, and both central and satellite galaxies exhibit inside-out quenching (independent of their classification).  In the case of main sequence and green valley galaxies, both central galaxies and satellite galaxies exhibit the same systematic trend in which the gradients of the \age{} profiles for the TNG100 galaxies is steeper than those of observed galaxies.  In the case of quenched galaxies, the gradients of the \age{} profiles for TNG100 central and satellite galaxies agree well with those of observed central and satellite galaxies.  As with the complete samples of galaxies, these results likely reflect differences in which ages of stellar populations were obtained for the observed vs.\ simulated galaxies.

\section{Summary \& Conclusions} \label{sec:conclusion}

We have investigated stellar ages and star formation rates in TNG100 galaxies on kiloparsec scales. We identified stellar particles that formed in the last 20 or 100~Myr and used a standard SPH-smoothing technique to map stellar mass density and resolved SFR into 2-d maps. From the maps of individual galaxies, we obtained simulated spaxels and used them to identify a resolved star formation main sequence ($\Sigma_* - \Sigma_{\mathrm{SFR}}$ relation). Allowing for differences in resolution and fitting techniques, we find that the global and resolved SFMS for TNG100 galaxies is in good agreement with previous results from observations and other simulations.

Further, we constructed radial profiles of a parameter, \SFR, that identifies the offset from the rSFMS based on the $\Sigma_*$ of a spaxel. Additionally, we constructed radial profiles of luminosity-weighted age, \age, by binning stellar particles in circular apertures and summing over the particles' birth time and luminosity. 
The main conclusions from the radial profiles are:

\begin{itemize}

    \item In agreement with previous studies, we see evidence for inside-out quenching in high-mass galaxies ($\log_{10} M_*/M_\odot > 10.5$, determined via the total mass of all bound stellar particles). 
    Main sequence and green valley galaxies with $\log_{10} M_*/M_\odot > 10.5$ have steeply falling \age{} profiles, indicating that there has been less recent star formation in the centers of these galaxies, with stellar particles in the centers of these galaxies having the oldest ages.

    \item In contrast, low-mass galaxies ($\log_{10} M_*/M_\odot < 10.5$ ) do not show evidence for inside-out quenching, having much flatter, slightly positive gradients in \age{} profiles (and corresponding negative gradients in \SFR{} profiles). 

    \item The \age{} profiles of central and satellite galaxies are remarkably similar. This is the case for the complete sample, and also the case when centrals and satellites are divided into low- and high-mass systems. Overall, we find that the mass of the galaxy is a greater determinant of profile shape than the galaxy's categorization as a central or satellite. For green valley galaxies, high-mass ($\log_{10} M_* / M_\odot >10.0$) centrals and satellites have \age{} profiles with steep negative gradients, while low-mass centrals and satellites have \age{} profiles with positive gradients. 

    \item TNG100 does not reproduce the \SFR{} and \age{} radial profiles of observed starburst galaxies, for which recent star formation has occurred in the centers of the galaxies. This discrepancy is likely attributable to insufficient spatial resolution in the simulation.
\end{itemize}

Our results for TNG100 main sequence, green valley, and quenched galaxies are in general agreement with previous results.  That is, high-mass galaxies appear to quench from the inside-out, while the quenching of low-mass galaxies is not driven from the center, even for galaxies that fall below the main sequence. That the simulation is able to reproduce trends seen in observed galaxies supports further studies of \age{} radial profiles for TNG100 galaxies. In future work, we will investigate the dependence of the \age{} radial profile on intrinsic and environmental parameters, such as dark matter halo mass, central black hole mass and accretion rate, cumulative AGN feedback energy imparted on the galaxy, and local galaxy overdensity. Such an investigation will constrain which parameters drive quenching in central and satellite galaxies.

\begin{acknowledgments}
The authors thank the anonymous reviewer for comments that improved the clarity of this manuscript. We would like to acknowledge the work and documentation provided by the IllustrisTNG team that has made this paper possible. The IllustrisTNG simulations were undertaken with compute time awarded by the Gauss Centre for Supercomputing (GCS) under GCS Large-Scale Projects GCS-ILLU and GCS-DWAR on the GCS share of the supercomputer Hazel Hen at the High Performance Computing Center Stuttgart (HLRS), as well as on the machines of the Max Planck Computing and Data Facility (MPCDF) in Garching, Germany. The additional computational work done for this paper was performed on the Shared Computing Cluster which is administered by Boston University’s Research Computing Services. This work was partially support by a Summer 2021 Massachusetts Space Grant Fellowship and NSF grant AST-2009397.
\end{acknowledgments}

%

\vspace{5mm}
\facilities{}


\software{Astropy \citep{astropy:2013, astropy:2018, astropy:2022}} 




\bibliography{bib}{}
\bibliographystyle{aasjournal}



\end{document}